\documentclass[10pt,aps,prl,twocolumn,,superscriptaddress]{revtex4-1}
\usepackage{amsmath}
\usepackage{graphicx}
\usepackage{verbatim}
\usepackage{color}   
\usepackage{dcolumn}
\usepackage{bm}
\begin{document}

\renewcommand{\figurename}{FIG.}

\title{Specific Heat Study of 1D and 2D Excitations in the Layered Frustrated Quantum Antiferromagnets Cs$_2$CuCl$_{4-x}$Br$_x$}

\author{U. Tutsch}
\affiliation{Physikalisches Institut, Goethe-Universit\"at, 60438 Frankfurt (M), Germany}
\author{O. Tsyplyatyev}
\affiliation{Institut f\"ur Theoretische Physik, Goethe-Universit\"at, 60438 Frankfurt (M), Germany}
\author{M. Kuhnt}
\affiliation{Physikalisches Institut, Goethe-Universit\"at, 60438 Frankfurt (M), Germany}
\author{L. Postulka}
\affiliation{Physikalisches Institut, Goethe-Universit\"at, 60438 Frankfurt (M), Germany}
\author{B. Wolf}
\affiliation{Physikalisches Institut, Goethe-Universit\"at, 60438 Frankfurt (M), Germany}
\author{P. T. Cong}
\affiliation{Physikalisches Institut, Goethe-Universit\"at, 60438 Frankfurt (M), Germany}
\author{F. Ritter}
\affiliation{Physikalisches Institut, Goethe-Universit\"at, 60438 Frankfurt (M), Germany}
\author{C. Krellner}
\affiliation{Physikalisches Institut, Goethe-Universit\"at, 60438 Frankfurt (M), Germany}
\author{W. A{\ss}mus}
\affiliation{Physikalisches Institut, Goethe-Universit\"at, 60438 Frankfurt (M), Germany}
\author{B. Schmidt}
\affiliation{Max-Planck-Institut f\"ur Chemische Physik fester Stoffe, 01187 Dresden, Germany}
\author{P. Thalmeier}
\affiliation{Max-Planck-Institut f\"ur Chemische Physik fester Stoffe, 01187 Dresden, Germany}
\author{P. Kopietz}
\affiliation{Institut f\"ur Theoretische Physik, Goethe-Universit\"at, 60438 Frankfurt (M), Germany}
\author{M. Lang}
\affiliation{Physikalisches Institut, Goethe-Universit\"at, 60438 Frankfurt (M), Germany}

\date{\today}

\begin{abstract}
We report an experimental and theoretical study of the low-temperature specific heat $C$ and magnetic susceptibility $\chi$ of the layered anisotropic triangular-lattice spin-1/2 Heisenberg antiferromagnets Cs$_2$CuCl$_{4-x}$Br$_x$ with $x$ = 0, 1, 2, and 4. We find that the ratio $J'/J$ of the exchange couplings ranges from 0.32 to $\approx 0.78$, implying a change (crossover or quantum phase transition) in the materials' magnetic properties from one-dimensional (1D) behavior for $J'/J < 0.6$ to two-dimensional (2D) behavior for $J'/J \approx 0.78$ behavior. For $J'/J < 0.6$, realized for $x$ = 0, 1, and 4, we find a magnetic contribution to the low-temperature specific heat, $C_{\rm m} \propto T$, consistent with spinon excitations in 1D spin-1/2 Heisenberg antiferromagnets. Remarkably, for $x$ = 2, where $J'/J \approx 0.78$ implies a 2D magnatic character, we also observe $C_{\rm m} \propto T$. This finding, which contrasts the prediction of $C_{\rm m} \propto T^2$ made by standard spin-wave theories, shows that Fermi-like statistics also plays a significant role for the magnetic excitations in frustrated spin-1/2 2D antiferromagnets.
\end{abstract}

\pacs{}

\maketitle

\textit{Introduction.}\ \textendash\ Spin-1/2 antiferromagnets on frustrated lattices are considered a source of intriguing phenomena. The interplay of geometric frustration and strong quantum fluctuations is known to weaken or even destroy magnetic order and may give rise to novel liquid-like states, so-called quantum spin liquids \cite{Bal10,Sav17}, or non-trivial quantum phase transitions, see, e.g.~\cite{Kog00}. The theoretical understanding of the phases involved and their experimental identification pose major challenges in current research on correlated quantum many-body systems. In this respect specific heat measurements play an important role \cite{Yam08,Yam11,Li15} for characterizing the nature of the excitations of these phases and for the determination of the entropy associated with them.

Layered spin-1/2 Heisenberg antiferromagnets with an anisotropic triangular arrangement of spins which interact by exchange coupling constants $J$ and $J'$ (see Fig.\,1), represent an interesting family of such correlated systems where the possibility of a frustration-induced quantum phase transition has been discussed \cite{Yun06,Hay07,Hei09}. Moreover, when close enough to the Mott metal-insulator transition \cite{Mot05,Sav17} where additional interactions, such as ring exchange, become relevant, these systems may also support a quantum spin liquid state. In triangular-lattice spin-1/2 Heisenberg antiferromagnets, the geometric frustration supports an effective decoupling of the spin chains, defined by the dominant coupling constant $J$, thus extending the range where 1D behavior dominates to relatively high $J'/J$ values. Whether the 2D state is reached by a crossover or by a quantum phase transition \cite{Yun06,Hei09,Reu11,Che13} is still under debate.

There is general consensus \cite{Yun06,Koh07,Hay07,Sta07,Hei09,Bal10} that for $J'/J < 0.6$ the 1D behavior prevails where fractionalized $S = 1/2$ spinon excitations with fermionic character propagate along the chains. As rigorously shown by Bethe-ansatz calculations \cite{Tak73,Klu98} for the antiferromagnetic Heisenberg chain, these spin-1/2 excitations are reflected in a low-temperature contribution to the magnetic specific heat, $C_{\rm m}$, which (in good approximation) varies linearly with temperature $T$, i.e., $C_{\rm m}\propto T$. On the other hand, for 2D quantum antiferromagnets, expected for $J'/J$ sufficiently larger than 0.6, the situation is less clear. At first glance this seems surprising considering that the specific heat is a sensitive quantity for probing the dimensionality of the low-lying excitations, as, e.g., verified for lattice excitations \cite{Her76,Hon00,Pop12}. The difficulty for 2D antiferromagnets lies in the quantum nature of $S=1/2$ spin operators which commute on different sites (like bosons) but locally, on the same site satisfy the $SU(2)$ algebra including a Fermi-like anticommutation relation between spin-ladder operators. Thus depending on the model applied, different results have been obtained for $C_{\rm m}(T)$. According to modified spin-wave \cite{Tak89} and Schwinger-boson-based mean-field \cite{Aue88} theory, a $C_{\rm m}\propto T^2$ behavior was proposed. This contrasts with $C_{\rm m}\propto T^{\nu}$ and $0.67\leq\nu\leq 1$, obtained using Resonating Valence Bond theory \cite{And87,And88}, Wigner-Jordan fermions \cite{Wan92}, Gutzwiller projection of fermionic mean-field states \cite{Mot05}, and a recent spin Hartree-Fock approach \cite{Wer18}.

\begin{figure}[b]
	\includegraphics[width=1.0\columnwidth]{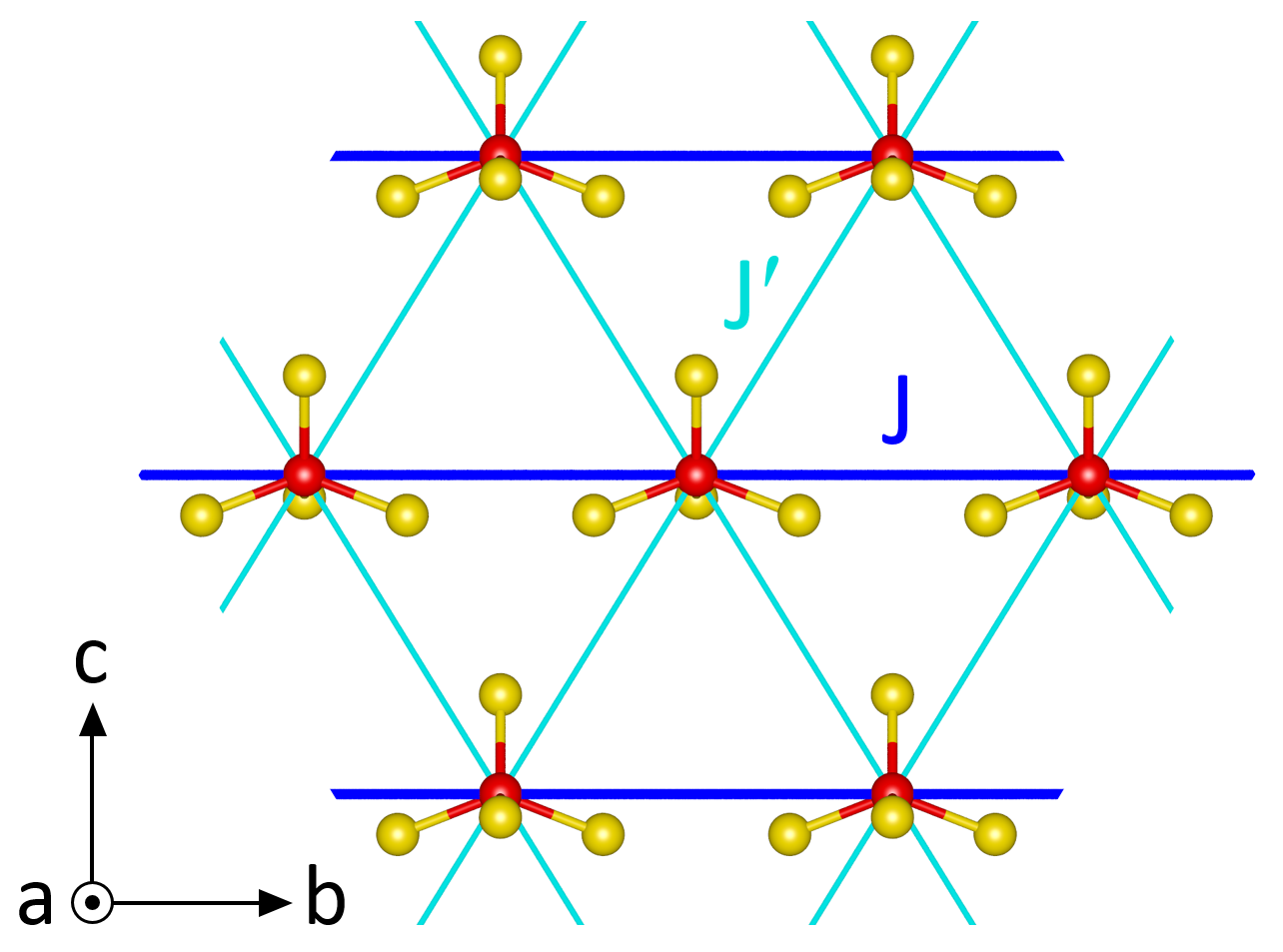}
	\caption{Spin-coupling scheme of the anisotropic triangular-lattice Heisenberg antiferromagnet as realized in Cs$_2$CuCl$_{4-x}$Br$_x$ where the $S=1/2$ spins of the Cu$^{2+}$ ions (red spheres) are tetrahedrally coordinated by the halide ions (yellow spheres).}
	\label{Fig1}
\end{figure}

Thus, low-temperature specific heat measurements on triangular-lattice Heisenberg antiferromagnets, covering the range from $J'/J < 0.6$ (1D) to $J'/J$ significantly above 0.6 (2D), are of great interest for identifying the character of the low-energy excitations in frustrated 2D antiferromagnets, thereby settling this fundamental issue. Here we report an experimental study of the low-temperature specific heat on Cs$_2$CuCl$_{4-x}$Br$_x$ single crystals with $x$ = 0, 1, 2, and 4 where $J'/J$ is found to span a wide range from $0.32$ to $\approx 0.78$. This system thus offers the possibility to study, on a series of isostructural compounds, the character of the low-lying excitations in the different regimes and to test the theoretical predictions.

\textit{The chosen quantum-spin system.}\ \textendash\ The two border compounds of the Cs$_2$CuCl$_{4-x}$Br$_x$ system ($x$ = 0 and 4), where Cu$^{2+}$ ions carry well-localized $S = 1/2$ spins, have been studied intensively for more than 15 years. A comprehensive characterization was provided by neutron-scattering experiments which revealed the geometry and size of the spin-spin interactions \cite{Col02,Col03,Ono03,Ono04}. According to these studies, both compounds are good realizations of a layered anisotropic triangular-lattice Heisenberg antiferromagnet. The dominant interaction $J$ runs along the $b$-direction thereby forming chains. These chains interact with each other via the weaker diagonal interaction $J'$ in the $bc$-plane, cf.\,Fig.\,1. For the two border compounds values of $J/k_{\rm B} = 4.34\rm\,K$, $J'/k_{\rm B} = 1.48\rm\,K$ (Cs$_2$CuCl$_4$) \cite{Col02} and $J/k_{\rm B} = 13.9\rm\,K$, $J'/k_{\rm B} = 6.49\rm\,K$ (Cs$_2$CuBr$_4$) \cite{Ono03} were found. Besides the dominant couplings $J$ and $J'$, a weak Dzyaloshinskii-Moriya interaction with components $D_a$, $D_c$ was observed ($D_a/k_{\rm B} = 0.23\rm\,K - 0.33\,K$, $D_c/k_{\rm B} = 0.36\rm\,K$ for $x = 0$) \cite{Col02,Fay13} along with a weak interlayer interaction $J_{\bot}$ ($J_{\bot}/k_{\rm B} = 0.20\rm\,K$ for $x = 0$ \cite{Col02} and $J_{\bot}/k_{\rm B} < 0.64\rm\,K$ for $x = 4$ \cite{Ono03}).

The small ratio $J_{\bot}/J$ for the $x$ = 0 and 4 compounds implies low-dimensional magnetic behavior over a wide range of temperatures $T\gg J_{\bot}/k_{\rm B}$. However, $J_{\bot}$ is still strong enough to generate 3D antiferromagnetic ordering with N\'{e}el temperatures $T_{\rm N}$ of 0.6\,K ($x$ = 0) \cite{Col96} and 1.4\,K ($x$ = 4) \cite{Ono03}. On the other hand, as shown in \cite{Ono05}, partial substitution of Cl for Br in Cs$_2$CuBr$_4$ can lower $T_{\rm N}$ significantly to at least 0.6\,K and possibly even further, suggesting that for certain Br concentrations $T_{\rm N}$ values much smaller than 0.6\,K might be possible. An ordering temperature as low as possible is desirable in order to search for signatures of genuine 1D or 2D behavior and its dependence on $J'/J$.

\textit{Crystals.}\ \textendash\ Single crystals of Cs$_2$CuCl$_{4-x}$Br$_x$ were grown from aqueous solutions at temperatures of about $50^{\circ}\rm C$ and then characterized by structural and energy-dispersive x-ray investigations, see Refs. \cite{Con11,SM}). Under these conditions, the substitution of Br for Cl (and vice versa) is site selective \cite{Kru10,Con11,Wel15}, an important aspect which ensures a well-ordered halide sublattice. Note that for crystals grown by the Bridgman method (as in Ref. \cite{Ono05}), the high temperatures of about $600^{\circ}\rm C$ used there imply a random distribution of Br and Cl on the halide sites. Thus, the growth from an aqueous solution provides crystals with a regular halide sublattice structure not only for the two border compounds ($x$ = 0, 4) but also for the two intermediate systems with $x$ = 1 and 2. For $x$ = 3 the site-selective occupation does not lead to a well-ordered halide sublattice as there are two Cl(3) sites in each copper-halide tetrahedron, both occupied by Br and Cl with equal probability.

\begin{figure}[t]
	\includegraphics[width=1.0\columnwidth]{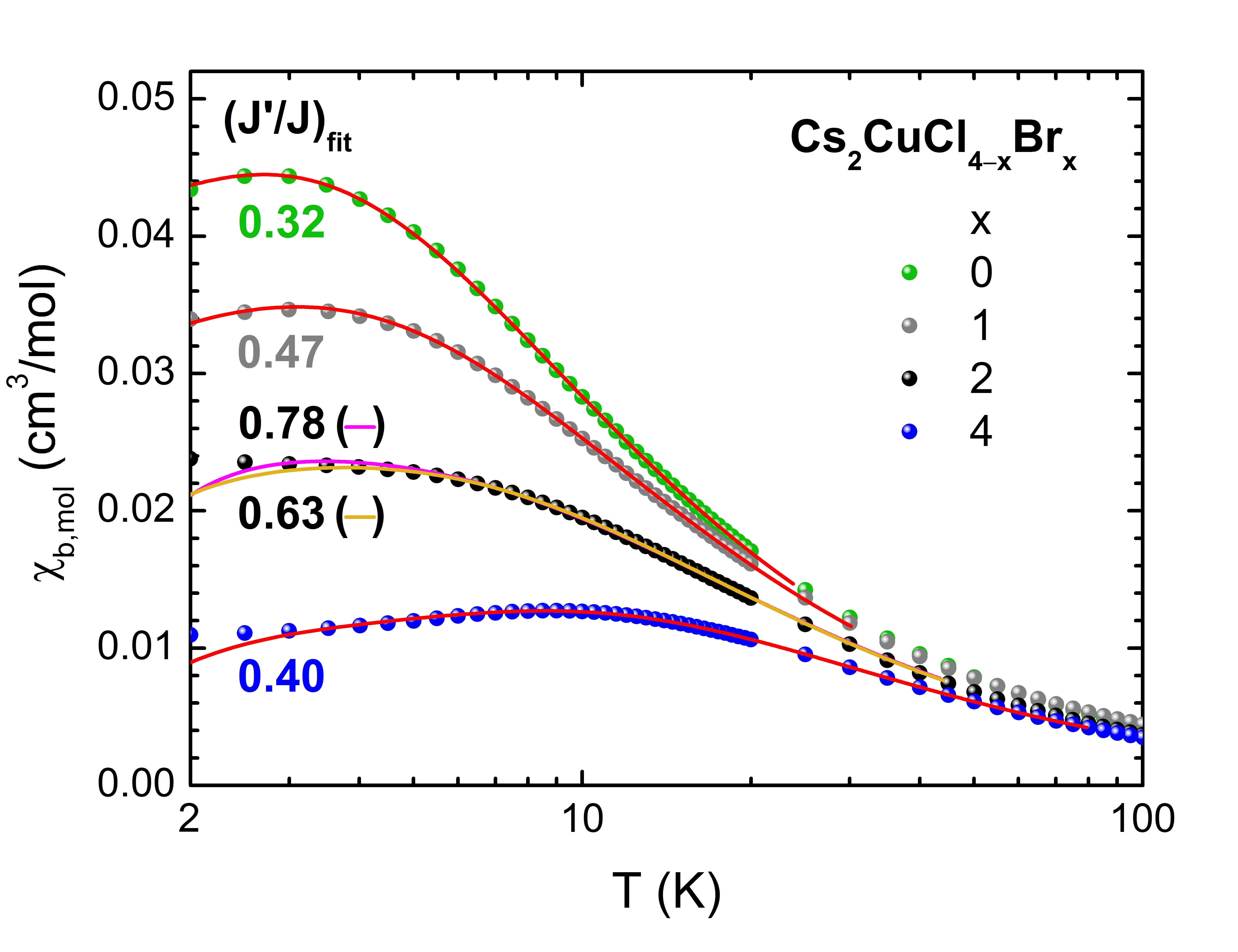}
	\caption{Magnetic susceptibility (in cgs units) measured along the $b$-axis of Cs$_2$CuCl$_{4-x}$Br$_x$ as function of temperature. Circles: experimental data; Lines: fit curves based on the model of the $S=1/2$ Heisenberg antiferromagnet with anisotropic triangular lattice using the finite-temperature Lanczos method. Colored numbers are $J'/J$ values obtained by this method.}
	\label{Fig2}
\end{figure}

\textit{Magnetic Susceptibility.}\ \textendash\ For characterizing the single crystals of Cs$_2$CuCl$_{4-x}$Br$_x$ used in our study and for determining their coupling constants $J$ and $J'$, we have measured the low-field ($\mu_0H = \rm0.1\,T$) molar magnetic susceptibility $\chi_{\rm mol} = (1/n)M/H \approx (1/n)\partial M/\partial H$ ($n$ being the amount of substance) in the temperature range $2\,{\rm K} \leq T \leq \rm100\,K$ for various Br concentrations $x$ covering the whole concentration range from $x$ = 0 to 4 \cite{Con11}. A Quantum Design superconducting quantum interference device magnetometer was used for this purpose. After correcting for the temperature-independent diamagnetic core contribution and the magnetic contribution of the sample holder, the data were analyzed by using theoretical calculations of $\chi(T)$ for the anisotropic triangular-lattice $S = 1/2$ Heisenberg antiferromagnet based on the finite-temperature Lanczos method \cite{Sch14,Sch15}. The $g$-factor and the coupling constants $J$ and $J'$ were used as free parameters in the fits. The so-obtained results are presented in Fig.\,2 for the Br concentrations $x$ = 0, 1, 2, and 4. The agreement between the model calculations and the experimental data is very good above the temperatures where the fit curves show their maxima. For temperatures significantly below the maximum, which is of relevance only for $x = 4$, the fits are slightly affected by finite-size effects leading to a rapid decrease of the calculated susceptibility for $T\to 0$. In case of the $x = 2$ compound, where the $\chi$ data lack a well-pronounced maximum \cite{Tha18}, fits of comparable quality could be obtained for $J'/J$ values ranging from 0.63 up to 0.78.

\textit{Specific heat.}\ \textendash\ The results for the magnetic specific heat $C_{\rm m}$ (divided by temperature) are shown in Fig.\,3 for single crystals of all compounds with a well-ordered halide sublattice ($x$ = 0, 1, 2, 4). Data were taken from 40\,mK to about 20\,K. The high-temperature data ($T > 1.8\rm\,K$) were obtained with a PPMS relaxation calorimeter (Quantum Design) whereas for the low-temperature range a self-constructed relaxation calorimeter adapted to a $^3$He-$^4$He dilution refrigerator was used ($x$ = 1, 2). In case of the two border compounds ($x$ = 0, 4) and low temperatures, data from literature \cite{Rad05,Ono05} were taken which overlap with our results in the temperature range from 1.8\,K to 6\,K. The agreement between both data sets is very good except for $x$ = 0 and $1.8\,{\rm K} \leq T \leq \rm3.0\,K$, where our data lie somewhat below (maximally 7\,\%) those of Ref. \cite{Rad05}. The magnetic specific heat $C_{\rm m}$ was obtained from the total specific heat $C$ by subtracting the nuclear contribution $C_{\rm n} = A/T^2$, caused by the hyperfine interaction of the copper ions, as well as the phonon contribution $C_{\rm ph}$. As described in detail in Ref. \cite{SM} and shown exemplarily for the $x$ = 2 compound (inset of Fig.\,3), the nuclear contribution becomes relevant only below about 100\,mK whereas the phonon contribution starts to become significant above about 2\,K.

\begin{figure}[b]
	\includegraphics[width=1.0\columnwidth]{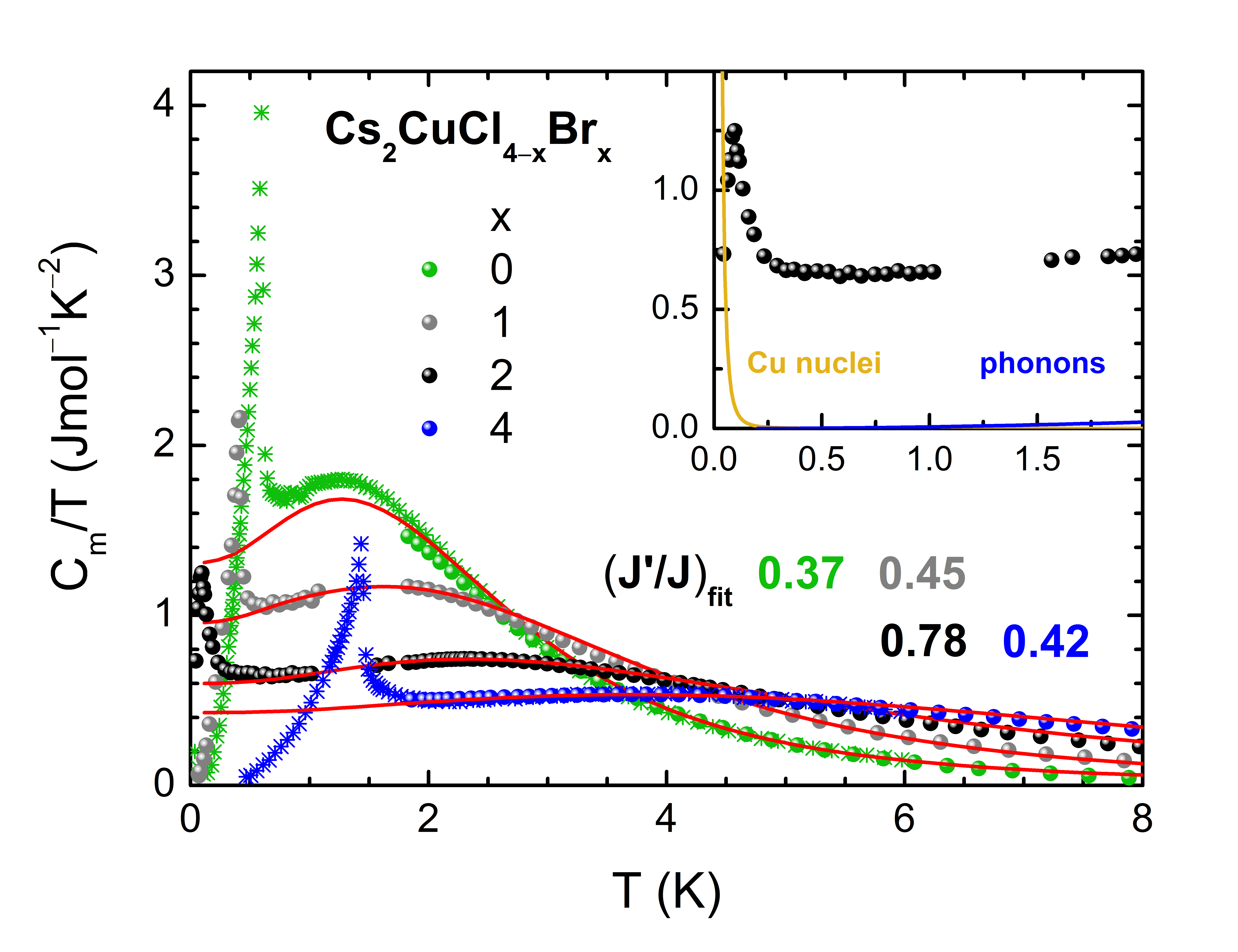}
	\caption{Magnetic specific heat devided by temperature $C_{\rm m}/T$ of Cs$_2$CuCl$_{4-x}$Br$_x$ as function of temperature. Circles (crosses): experimental data of this work (from reference \cite{Rad05} for $x=0$ and \cite{Ono05} for $x=4$); Lines: fit curves based on the model of the $S=1/2$ Heisenberg antiferromagnet with anisotropic triangular lattice using the spin Hartree-Fock approach. The colored numbers are the $J'/J$ values which have been obtained by this method; Inset: $C_{\rm m}/T$ of Cs$_2$CuCl$_2$Br$_2$ below 2\,K. Yellow (blue) line: contribution from the copper nuclei (phonons).}
	\label{Fig3}
\end{figure}

The data in Fig.\,3 represent the central result of this study and contain three important pieces of information. Besides the identification of phase transition anomalies for the recently discovered intermediate compounds $x$ = 1 and 2, the data can be used for an independent determination of the $J'/J$ values for all crystals under investigation. In addition, and of particular interest here is the determination of the temperature dependence of $C_{\rm m}$, i.e., the identification of a potential power-law behavior at low temperatures $T < T_{\rm max}$, with $T_{\rm max}$ the temperature where $C_{\rm m}$ adopts a broad maximum.

Figure 3 reveals clear evidence for phase transition anomalies in $C_{\rm m}/T$ also for the intermediate compounds with $x = 1$ and $2$. In analogy to the border compounds $x = 0$ and $4$, we assign these transitions to the onset of long-range antiferromagnetic order at $T_{\rm N} = 0.41\rm\,K$ ($x$ = 1) and 0.095\,K ($x$ = 2). Remarkably, for the $x$ = 2 compound, $T_{\rm N}$ is strongly suppressed as compared to the other compounds, reflecting a particularly high degree of frustration for this material. For a quantitative determination of the $J'/J$ values from the data in Fig.\,3, we performed model calculations for $C_{\rm m}$. Instead of using the Lanczos method, which becomes too inaccurate at such low temperatures ($T \ll T_{\rm max}$), we use the recently proposed spin Hartree-Fock approach \cite{Wer18}. This new method has been successfully applied in Ref.\,\cite{Wer18} to the case of an antiferromagnetic Heisenberg chain, for which the specific heat is known with high accuracy \cite{Klu98}. The spin Hartree-Fock approach well reproduces the Bethe-ansatz results at $T < 0.9\,T_{\rm max}$, but yields a slightly too high (15\%) value for $C_{\rm m}$ around $T_{\rm max}$. However, as we apply this method only for fitting $C_{\rm m}(T)$ up to about $0.7\,T_\textrm{max}$, this limitation is of no relevance here.

\begin{figure}[b]
	\includegraphics[width=1.0\columnwidth]{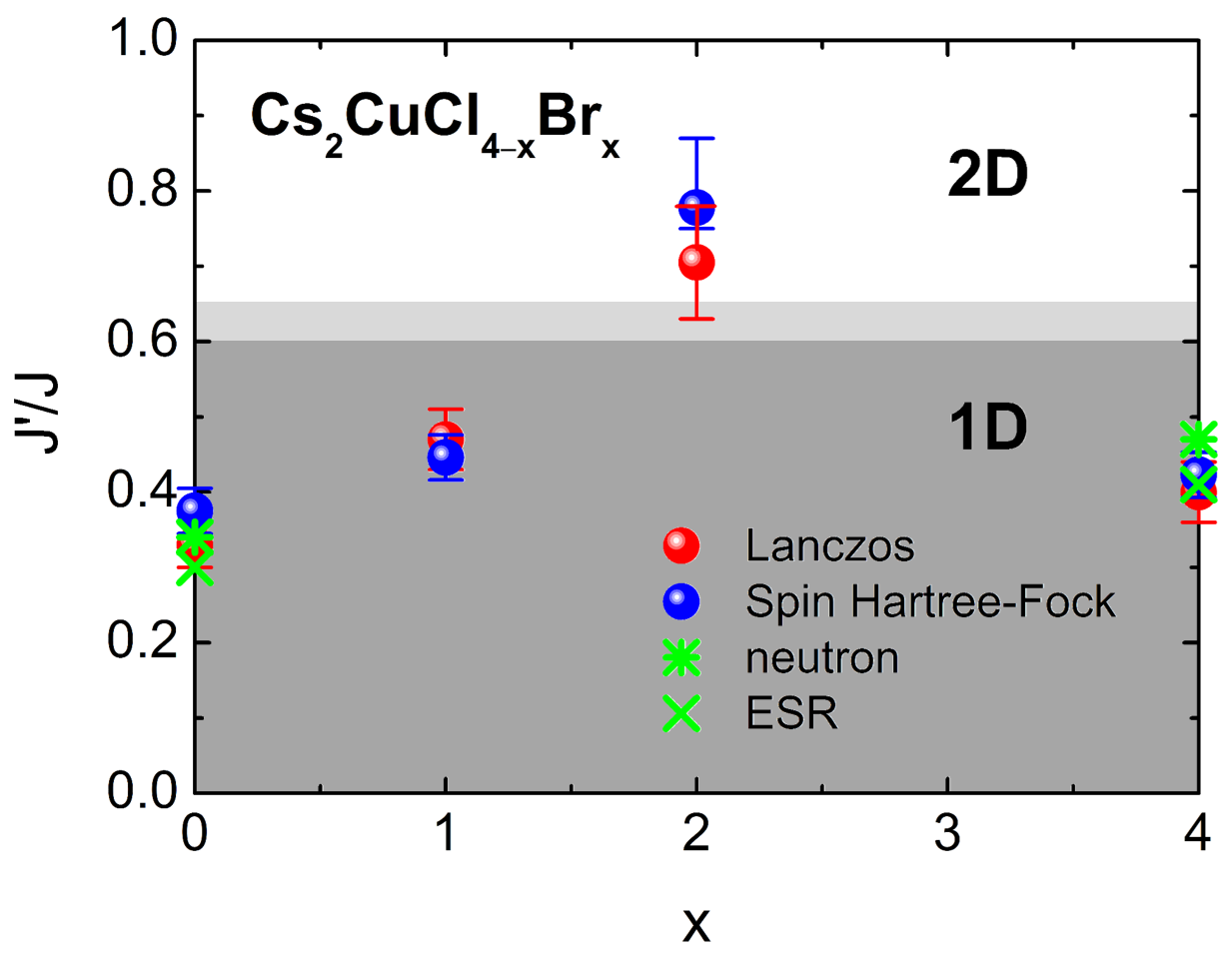}
	\caption{Coupling ratio $J'/J$ of Cs$_2$CuCl$_{4-x}$Br$_x$ as function of the Br concentration $x$. Red (blue) spheres: values obtained with the Lanczos method (spin Hartree-Fock approach) using the susceptibility (magnetic specific heat) data; green crosses: literature data obtained by neutron scattering \cite{Col02,Col03} or ESR \cite{Zvy14} experiments. At a critical value of $J'/J$, which is close to 0.6, a crossover or quantum phase transition from 1D to 2D magnetic behavior is expected.}
	\label{Fig4}
\end{figure}

The so-derived fit curves for the specific heat, included as red lines in Fig.\,3, provide an excellent description of the experimental data above $T_{\rm N}$ in the temperature range shown in Fig.\,3. In addition, the $J'/J$ values obtained from the least-square fits are very close to those revealed by the Lanczos fits to the susceptibility data (see Fig.~2). For a compilation of the $J'/J$ values of all compounds studied see Fig.~4. In addition Fig.\,4 demonstrates that for $x = 0$ and $4$ the $J'/J$ values derived from the spin Hartree-Fock description are also in good agreement with the values obtained by neutron scattering \cite{Col02,Col03} and ESR \cite{Zvy14} studies. For $x$ = 2, however, the spin Hartree-Fock result yields $J'/J = 0.78_{-0.03}^{+0.09}$ which lies at the upper end of the range $J'/J = 0.63 - 0.78$ obtained from the least-square Lanczos fits. In view of the considerable uncertainties involved in pinpointing the $J'/J$ value for $x$ = 2 from fits to the $\chi(T)$ data, and given the fact that a description of the specific heat for this compound with values $J'/J < 0.75$ is of distinctly less quality \cite{SM}, we consider $J'/J = 0.78_{-0.03}^{+0.09}$ to be reliable.

\textit{Discussion.}\ \textendash\ In discussing the low-temperature specific heat data with regard to potential power-law behavior, it is obvious that the occurrence of a phase transition at $T_{\rm N}$ due to a weak interlayer coupling $J_{\bot}$ imposes some limitations. These restrictions are more severe for the $x=0$ compound but of less relevance for $x = 1$, 2, and 4. At the same time, as the model calculations based on the spin Hartree-Fock approach provide a very good description of $C_{\rm m}(T)$ for $T_{\rm N}\leq T\lesssim 0.7\,T_{\rm max}$, we can include these theoretical results in the discussion.

First, we focus on the compounds with $J'/J < 0.6$, realized for $x$ = 0, 1, and 4 with $J'/J = 0.37$, $0.45$, and $0.42$, respectively. For these compounds we find a low-temperature specific heat which approaches a $C_{\rm m}/T = const.$ behavior for $T\to 0$. This observation is consistent with the 1D magnetic behavior expected for $J'/J < 0.6$ \cite{Yun06,Koh07,Hay07,Sta07,Hei09,Bal10}. Remarkably, as shown in the inset of Fig.\,3 on expanded scales, a $C_{\rm m}/T = const.$ behavior over a rather wide temperature range is also revealed for the $x$ = 2 compound, which is already far inside the 2D regime. This experimental finding clearly contradicts the predicition of $C_{\rm m}/T \propto T$ made by the spin-wave theories based on bosons \cite{Aue88,Tak89}. Such a behavior is counterintuitive at first sight. However, it can be rationalized by considering the non-trivial statistics of magnetic excitations in the fully quantum case of spin-1/2, which are usually highly entangled states in low-dimensional quantum magnets involving many spins. Note that on different lattice sites $i\neq j$ the spin-1/2 operators commute, \emph{e.g.}, $S_{i}^{\alpha}S_{j}^{\beta}-S_{j}^{\beta}S_{i}^{\alpha}=0$, like bosons, but on the same lattice site the spin ladder operators $S_{i}^{\pm}$ obey the anti-commutation relation, $S_{i}^{-}S_{i}^{+}+S_{i}^{+}S_{i}^{-} =1$.
Thus, the many-body states consisting of a great number of magnetic excitations have a structure that is neither entirely symmetric with respected to permutation of two particles (Bose-like) nor entirely anti-symmetric (Fermi-like). The $C_{\rm m}/T \approx const.$ behavior observed in this experiment for a 2D antiferromagnet can be interpreted as a consequence of the Fermi-like part of the statistics (see details in \cite{SM}) which is taken into account by the theories in Refs. \cite{And87,And88,Wan92,Mot05,Wer18}.

In the 2D regime, realized in the $x$ = 2 compound, the frustration effects of a triangular lattice are generally expected to be strong since the antiferromagnetic couplings to the neighboring spins cannot be satisfied simultaneously resulting in a macroscopically degenerate ground state. Using the standard thermodynamic relation ($dS = C\,dT/T$) between the change of the entropy, $dS$, and the heat capacity, we can estimate the change of the total entropy $\Delta S = S(T=\infty)-S(T=0)$ by integrating the experimental $C_{\rm m}/T$ data from $T_1 = 40\rm\,mK$ up to $T\leq T_2 = 20\rm\,K$ and then extrapolating the so obtained $\Delta S(T) = S(T) - S(T_1)$ to $T = \infty$ and $T_1 = 0$. We find $\Delta S = (0.95\pm 0.05) R\ln2$ with $R$ the gas constant (see Ref. \cite{SM} for details) which is almost identical to the full entropy of $R\ln2$ expected for $S = 1/2$ spins. This result implies an upper bound of $S(0)\leq 0.05\,R\ln2$ for the residual entropy of the ground state. It is interesting to compare this value with the well-established result for the isotropic triangular-lattice Ising model yielding a residual entropy of $S(0) = 0.34\,R$ \cite{Wan50}. This marked difference in the geometrical frustration can be attributed to the dominant role of quantum fluctuations in spin-1/2 Heisenberg systems as opposed to Ising systems. The quantum uncertainty of a spin-1/2 is of the order of its size and can result in a fully non-degenerate ground state.

We note that layered triangular-lattice spin-1/2 systems with similar ratios $J'/J\approx 0.74$ to 0.84 are realized in the organic charge-transfer salts $\kappa$-(BEDT-TTF)$_2$Cu$_2$(CN)$_3$ and EtMe$_3$Sb[Pd(dmit)$_2$]$_2$. These systems, where the low-temperature specific heat also varies linearly in $T$ \cite{Yam08,Yam11}, are considered as prime candidates for a quantum spin liquid \cite{Shi03,Ito08}. In contrast to the present Cs$_2$CuCl$_2$Br$_2$ compound, described by well-localized spins, these organic materials are located rather close to the Mott transition so that a description based on a pure spin Hamiltonian appears inappropriate \cite{Sav17,Mot05}.

\textit{Conclusions.} \textendash\ Measurements of the low-temperature specific heat have been performed on four members of the layered anisotropic triangular-lattice spin-1/2 Heisenberg antiferromagnets Cs$_2$CuCl$_{4-x}$Br$_x$ all of which show a structurally well-ordered halide sublattice. The materials span a wide range of the ratio of coupling constants $0.32\leq J'/J\lesssim 0.78$, implying a change from 1D magnetic behavior for $J'/J < 0.6$ ($x$ = 0, 1, 4) to 2D behavior for $J'/J \approx 0.78$ ($x = 2$). Our central finding is that for the frustrated 2D case, the magnetic specific heat varies linearly in temperature, $C_{\rm m}\propto T$, reflecting a significant role of Fermi-like statistics in this 2D quantum antiferromagnet. Moreover, at variance to a naive expectation for such a strongly frustrated system, no indication of residual entropy is found within the experimental uncertainty. This observation, which is in marked contrast to the triangular-lattice Ising model, is attributed to the importance of quantum fluctuations in low-dimensional spin-1/2 Heisenberg systems.

This work was supported by the Deutsche Forschungsgemeinschaft (DFG) via the SFB/TR\,49. We thank Natalija van Well for her help in the single crystal growth.

\bibliography{Tutsch_Cs2CuCl4-xBrx_Lit}

\end{document}